\documentclass{IEEEtran}
\usepackage{amssymb}
\usepackage{amsfonts}
\usepackage{amsmath}
\usepackage{graphicx,cite,multicol,multirow,diagbox,booktabs,array}
\usepackage{subfig}
\usepackage{verbatim}
\usepackage{algorithm}
\usepackage{algorithmic}
\usepackage[justification=centering]{caption}
\usepackage[dvips]{color}
\usepackage{float}
\usepackage{rotating}

\usepackage{etoolbox}
\pretocmd{\maketitle}{%
  \markboth{The definitive version was published in in IEEE Transactions on Vehicular Technology, vol. 71, no. 1, pp. 1072-1076, Jan. 2022, doi: 10.1109/TVT.2021.3125361.}{}%
}{}{}

\ifCLASSOPTIONcompsoc
\fi
\ifCLASSINFOpdf
\else
\fi
\begin{document}
\title{Communication-efficient Coordinated  RSS-based Distributed Passive Localization via Drone Cluster}

\author{Xin Cheng,~Weiping Shi,~Wenlong Cai,~Weiqiang Zhu,~Tong Shen,~Feng Shu,~and~Jiangzhou Wang,~\IEEEmembership{Fellow,~IEEE}
\thanks{Xin Cheng,~Weiping Shi,~Tong Shen  are with School of Electronic and Optical Engineering, Nanjing University of Science and Technology, Nanjing, 210094, China. Email: xincstar23@163.com.}
\thanks{Weiqiang Zhu is with the 8511 Research Institute, China Aerospace Science and Industry Corporation, Nanjing, 210007, China.}
\thanks{Wenlong Cai is with the National Key Laboratory of Science and Technology on Aerospace Intelligence Control, Beijing Aerospace Automatic Control Institute, Beijing 100854, China.}
\thanks{Feng Shu is with the School of Information and Communication Engineering,  Hainan University,~Haikou,~570228, China and with the School of Electronic and Optical Engineering, Nanjing University of Science and Technology, Nanjing, 210094, China.}
 \thanks{Jiangzhou Wang is with the School of Engineering and Digital Arts, University of Kent, Canterbury CT2 7NT, U.K.}
}

\maketitle
\begin{abstract}
  Multi-UAV passive localization via received signal strength (RSS) is extremely important for wide applications such as rescue and battlefield combat. However, the energy consumption of UAVs is a key issue in this UAVs-enabled application. Usually, the communication overhead plays an important role in the energy consumption. To address this problem, we design two distributed methods for this multi-UAV system with considerable performance under low communication overhead. Firstly, a distributed majorize-minimization (DMM) method is proposed. To accelerate its convergence, a tight upper bound of the objective function from the primary one is derived.  Furthermore, a distributed estimation scheme using the Fisher information matrix (DEF)  is presented, only requiring one round of communication between edge UAVs and central UAV. Simulation results show that the proposed DMM outperforms the existing distributed iterative methods in terms of root of mean square error (RMSE) under low communication overhead. Moreover, the most communication-effective DEF with local search estimation performs much better than the proposed DMM in terms of RMSE, but has a higher computational complexity.
\end{abstract}


\IEEEpeerreviewmaketitle

\section{Introduction}
Unmanned Aerial Vehicle (UAVs) equipped with wireless communication modules and appropriate sensors have allowed appreciable paradigms in civilian and military applications. The wireless sensor network (WSN) formed by the cooperation between multiple UAVs can be the key support to applications and terrestrial networks, users, and communicating entities\cite{202177,9165125}.  As one of UAV-enabled applications, passive localization through multiple UAVs has been active in wide applications\cite{2021727,6735682,9360612,8960453}, such as rescue and battlefield combat. Compared to traditional localization paradigms, such as cellular localization and satellite localization, the UAV-enabled localization has distinctive advantages. Rapid deployment, flexible relocation and high chances of experiencing line-of-sight propagation path features have been perceived as promising opportunities to provide difficult service. Besides, the localization uncertainty can be eliminated effectively via corresponding trajectory planning of UAVs\cite{8960453}. The passive location techniques use the location related measurements of the radio frequency signals from the target. The popular types of techniques are mainly divided into time of arrival (TOA)\cite{6184254}, direction of arrival (DOA)\cite{8290952} and received signal strength (RSS)\cite{1247811}. Although less accuracy, the RSS-based localization is cost-effective and can be conveniently implemented, which fit the demands of UAV-enabled application well.

One of the main challenges faced by the multi-UAV passive RSS-based localization is the energy consumption.  As widely acknowledged, the energy consumption of UAVs is a key issue in UAVs-enabled applications\cite{2021772}.  For traditional centralized location methods, all  measurements gleaned from UAVs have to be sent to a centralized processor, then the centralized processor has to compute based on the mass received data. This process requires high communication overhead and computing cost, resulting in  huge energy consumption. Distributed methods may mitigate the above challenges dramatically by avoiding sharing local measurements and running the computation over the whole UAV-enabled network.  In \cite{6735872,6158612}, distributed gradient-type methods with different step sizes were proposed. Based on Taylor approximation, a distributed Guass-Newton method (DGN) for position estimation was studied in \cite{4291880,7737071}. In \cite{2018Parallel}, the authors delved into distributed successive convex approximation methods (DSCA) to solve the problems in several engineering fields,  including localization.  However the iterating rounds may be large, causing a large communication overhead. In fact, the communication overhead plays an important role in the energy consumption. Moreover, more transmission traffic increases the probability of being detected by a malicious user due to the open and broadcasting property of wireless transmission between central UAV and edge UAVs.

In this paper, we focus on designing distributed methods for multi-UAV passive RSS-based localization with considerable performance under low communication overhead. For the non-convex and non-linear estimation problem, we propose a distributed iterative method based on majorize-minimization with fast convergence, which is named DMM. Moreover, inspired by the work in the \cite{4838824},  a novel weight-based method with a better performance is proposed, named DEF, which is  extremely communication-efficient. Our main contribution is summarized as follows:
\begin{enumerate}
\item  In our model, drone group and trajectory knowledge are utilized to locate a fixed target. To exploit a distributed property of a drone cluster, a distributed majorize-minimization (DMM) method is proposed to solve RSS localization problem.  By finding an tight upper bound of the objective function approximating from the primary one, the proposed algorithm converges quickly, at most four iterations. Numeral results also show that the proposed DMM  has better performance in both computational complexity and accuracy than the existing distributed iterating methods under low communication overhead.
\item To dramatically reduce the communication overhead between central UAV and edge UAVs, a distributed estimation scheme utilizing the Fisher information matrix, called DEF, is proposed  with a local search as initialization. In this scheme, only one round of communication between central UAV and edge UAVs  is required. Simulation results show that the proposed DEF makes a significant RMSE performance improvement over the proposed DMM at the expense of a higher computational complexity.  Finally, the CRLB of distributed multi-UAV localization utilizing trajectory is  derived as a performance benchmark.
\end{enumerate}


\emph{Notations:} Boldface lower case and upper case letters denote vectors and matrices, respectively. Sign $(\cdot)^{T}$ and $|| \cdot ||$ denotes  transpose operation and norm. $\mathbb{E}\{\cdot\}$ represents expectation operation. $\mathrm{Tr}(\cdot)$ represents the trace of matrix.
\section{System Model}
\begin{figure}
  \centering
  \includegraphics[width=0.45\textwidth]{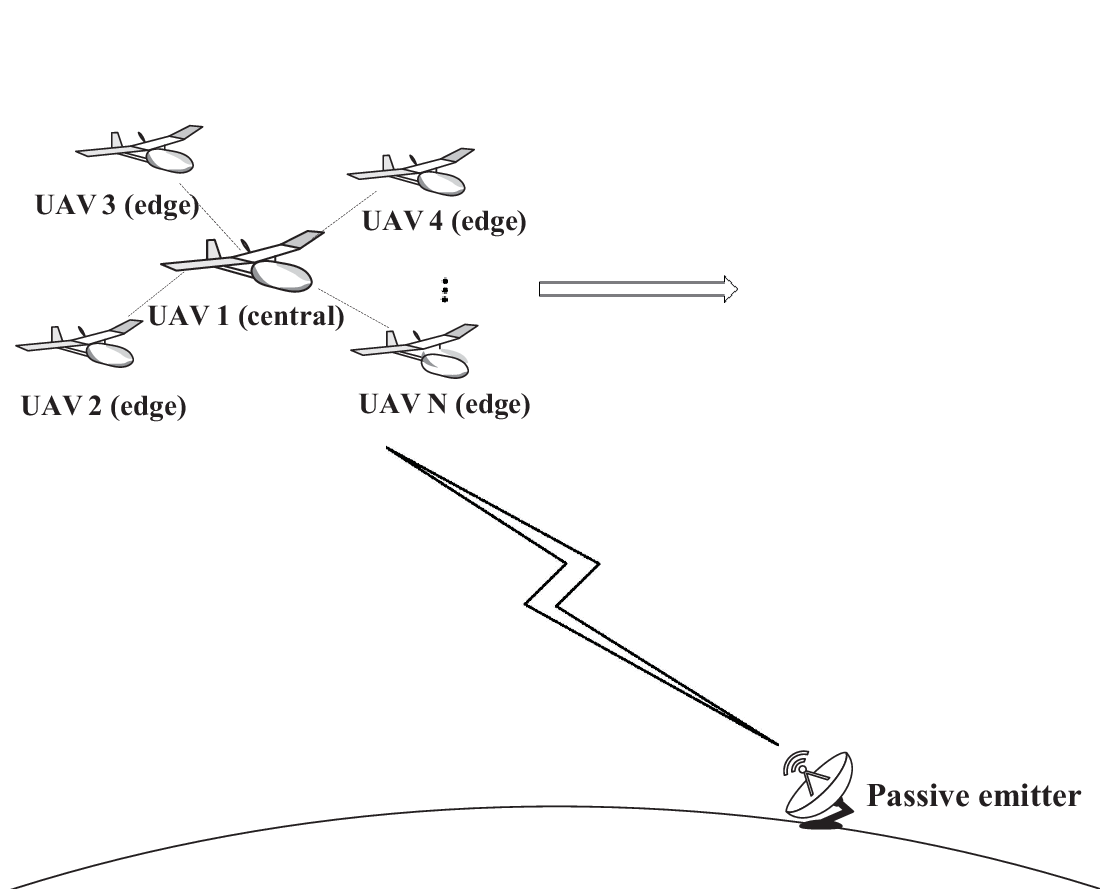}\\
  \caption{Multi-UAV passive localization system.}\label{sys}
\end{figure}
As illustrated in Fig.\ref{sys}, $N$ UAVs cooperate to  compute a single RF emitter in the area of interest (AOI).  Among  $N$ UAVs, there is one central UAV, who works as a  center of aggregating and processing  information from  other  UAVs. The central UAV's serial number is $1$  while the edge UAVs' serial numbers are from $2$ to $N$. Flying along the prearranged trajectory, each  UAV independently measures the RSS  from the target. The position of passive emitter is denoted as $\mathbf{s}=(x,y,z)$.  The $i$-th edge UAV measures $M_{i}$ times along its  trajectory.  The position of the $i$-th UAV at the $j$-th measurement is denoted as $\mathbf{u}_{i,j}=(x_{i,j},y_{i,j},z_{i,j})~1\leq i\leq N, 1\leq j\leq M_{i}$. Each UAV is aware of its own geographical position using the global positioning system (GPS).

The distance from   the  $j$-th measurement time of UAV  $i$  to the target can be expressed as
\begin{align}
d_{i,j}=\sqrt{(x-x_{i,j})^2+(y-y_{i,j})^2+(z-z_{i,j})^2}, \nonumber\\
1\leq i\leq N,~~1\leq j\leq M_{i}.
\end{align}

According to radio propagation path loss model (in decibels) \cite{1247811}, the received power of the  $j$-th measurement time of UAV  $i$ can be written as
\begin{align}\label{RSS}
&P_{i,j}=f_{i,j}(\mathbf{s})+\eta_{i,j},~f_{i,j}(\mathbf{s})\triangleq P_{0}-10\gamma_{i}\lg(\frac{d_{i,j}}{d_{0}}).
\end{align}
where $P_{0}$ denotes the power of source at reference distance $d_{0}$. $\gamma_{i}$ is the path loss exponent (PLE). The log-normal shadowing  term $\eta_{i,j}$ is a Gaussian random variable with mean zero and variance $\sigma_{i,j}$.

Stacking all measurement values of UAV  $i$ forms a $M_{i}$-dimensional column vector as follows
\begin{align}\label{RSS}
\mathbf{P}_{i}=\mathbf{f}_{i}(\mathbf{s})+\mathbf{\eta}_{i}.
\end{align}
which  yields the probability density function (pdf)
{\begin{align}\label{localML}
Q_{i}(\mathbf{s})=\frac{1}{(2\pi)^{\frac{M_{i}}{2}}|\mathbf{N}_{i}|}\exp\{(\mathbf{P}_{i}-\mathbf{f}_{i}(\mathbf{s}))^{T}\mathbf{N}_{i}^{-1}(\mathbf{P}_{i}-\mathbf{f}_{i}(\mathbf{s}))\},\nonumber\\
\end{align}
where $\mathbf{N}_{i}=\sigma_{i}\mathbf{I}_{M_{i}\times M_{i}}$ is the covariance matrix of $\mathbf{\eta}_{i}$.

Assuming measurement noises of different UAVs are dependent, the joint pdf of all UAVs is given by
\begin{small}
\begin{align}\label{BigML}
Q(\mathbf{s})=&\frac{1}{(2\pi)^{\frac{\sum_{i=1}^{N}M_{i}}{2}}\prod_{i=1}^{N}|\mathbf{N}_{i}|}   \nonumber\\ &\exp\{\sum_{i=1}^{N}(\mathbf{P}_{i}-\mathbf{f}_{i}(\mathbf{s}))^{T}\mathbf{N}_{i}^{-1}(\mathbf{P}_{i}-\mathbf{f}_{i}(\mathbf{s}))\}.
\end{align}
\end{small}
The centralized maximum likelihood of $\mathbf{s}$ is in the form of
\begin{align}\label{ml}
&\max_{\mathbf{s}}~~Q(\mathbf{s}) \nonumber\\
&\text{s. t.}~~~\mathbf{s}\in \mathrm{A},
\end{align}
where  $\mathrm{A}$ represents the area of interest.  However, this approach requires all edge UAVs to transmit measurement data to the central UAV, generating a heavy communication overhead.

\section{Proposed communication-efficiency distributed localization methods}\
To address the  problem of heavy communication overhead from centralized approach, the design of high-communication-efficiency distributed  methods is very important. In this section, two distributed communication-efficiency  methods are proposed as follows: DMM and DEF. Finally, the communication overhead and computational complexity are analysed.

\subsection{Proposed DMM}
In this subsection, we solve the RSS localization problem by MM algorithm in a distributed way. MM algorithm is a problem-driven algorithm that can take advantage of the problem structure. The key step to MM is constructing a surrogate function. By choosing surrogate function based on second order Taylor expansion, the MM can be implemented separably. In this way, the objective function at the $k$-th iteration in (\ref{BigML}) can be upper bounded as
\begin{align}\label{talor}
Q(\mathbf{s})\leq Q(\mathbf{s}^{k})+\mathbf{b}^{kT}(\mathbf{s}-\mathbf{s}^{k})+\frac{1}{2}(\mathbf{s}-\mathbf{s}^{k})^{T}\mathbf{M}(\mathbf{s}-\mathbf{s}^{k}),
\end{align}
where  $\mathbf{s}^{k}$ is a constant and  $\mathbf{b}^{k}=\nabla Q(\mathbf{s}^{k})$ stands for the corresponding gradient vector of objective $Q(\mathbf{s})$. Matrix $\mathbf{M}$ should satisfy the condition $\mathbf{M}\geq \nabla^{2}Q(\mathbf{s}), \forall  \mathbf{s} \in \mathrm{A}$.

Clearly, for a sufficiently large value of $\mathbf{M}$  large enough, the condition is naturally satisfied. But the corresponding updating size at each iteration is small, casing a huge number of iterations. In order to find a proper $\mathbf{M}$, we make an approximation to (\ref{RSS}) as follows
\begin{align}\label{approximation}
\tilde{d}_{i,j}\approx||\mathbf{u}_{i,j}-\mathbf{s}||_{2},
\end{align}
where $\tilde{d}_{i,j}=10^{\frac{P_{0}-P_{i,j}}{10\gamma_{i}}}d_{0}$. According to the least-squares (LS) criterion,  the optimization  problem in (\ref{BigML}) is reformulated as
\begin{align}\label{approRSSmodel}
\mathop{\min}_{\mathbf{s}} Q(\mathbf{s}) =\sum_{i=1}^{N}\sum_{j=1}^{M_{i}}(\tilde{d}_{i,j}-||\mathbf{u}_{i,j}-\mathbf{s}||_{2})^2.
\end{align}
After  calculation, we have
\begin{small}
\begin{align}
\nabla^{2}Q(\mathbf{s})
=&2\sum_{i=1}^{N}\sum_{j=1}^{M_{i}} (1-\frac{\tilde{d}_{i,j}}{||\mathbf{u}_{i,j}-\mathbf{s}||_{2}})\mathbf{I}\\\nonumber
&+\frac{\tilde{d}_{i,j}}{||\mathbf{u}_{i,j}-\mathbf{s}||_{2}^{3}}[(\mathbf{s}-\mathbf{u}_{i,j})(\mathbf{s}-\mathbf{u}_{i,j})^{T}],
\end{align}
\end{small}
and
\begin{align}
\mathbf{b}_{i}^{k}=\sum_{j=1}^{M_{i}}2(||\mathbf{u}_{i,j}-\mathbf{s}^{k}||_{2}-\tilde{d}_{i,j})\frac{\mathbf{s}^{k}-\mathbf{u}_{i,j}}{||\mathbf{s}^{k}-\mathbf{u}_{i,j}||_{2}},
\end{align}
where $k$ represents the $k$-th iteration and $\mathbf{s}^{k}$ is the initial value corresponding the $k$-th iteration. Let us define matrix $\mathbf{M}$ as follows
\begin{align}
\mathbf{M}=\lambda_{max}(\mathbf{\nabla^{2}Q(\mathbf{s})})\mathbf{I}=2K\mathbf{I},
\end{align}
where $K=\sum_{i=1}^{N}M_{i}$.  It is noted that $\mathbf{M}$ is only related to the total measurement times$\footnote{Notice that the $\mathbf{M}$ introduced by  $\lambda_{max}(\mathbf{\nabla^{2}Q(\mathbf{s})})\mathbf{I}$ with $Q(\mathbf{s})$ from (\ref{BigML}) is extremely relax.}$.

Finally, we make a summary of the procedure of the DMM algorithm. This method is composed of local updating step and fusion step. In the $(k+1)$-th iteration,  UAVs update locally based on $\mathbf{s}_{c}^{k}$ from the central UAV $1$ given by
\begin{align}\label{local}
\mathbf{s}_{i}^{k+1}=\mathbf{s}_{c}^{k}-\frac{N}{2K}\mathbf{b}_{i}^{k}.\\\nonumber
\end{align}
Then, the central UAV $1$ collects all $\mathbf{s}_{i}^{k+1}$ from edge UAVs to generate the new value as follows
\begin{align}\label{fusion}
\mathbf{s}_{c}^{k+1}=\frac{1}{N}(\mathbf{s}_{1}^{k+1}+\sum_{i=2}^{N}\mathbf{s}_{i}^{k+1}),
\end{align}
which should be distributed to all UAVs. Repeat the above process until the terminal criterion is reached.

\subsection{Proposed DEF}
To further reduce the communication overhead, we extend the distributed estimation scheme in \cite{4838824} to the multi-UAV localization with a completely different weight coefficient. In our scheme, each edge UAV only needs to send local estimation and the weighting coefficient to the central UAV $1$. The central UAV fuses the local estimates to produce the final result. Here, the reliable grid search with length size $\Delta d$ is selected as the local solver.

In the local estimation stage, the estimation of the $i$-th UAV, denoted as $\hat{\mathbf{s}}_{i}$, is  from its own measurements.  The estimation is treated  as an observation of the true position
\begin{small}
\begin{align}\label{localites}
\hat{\mathbf{s}}_{i}=\mathbf{s}+\mathbf{\xi}_{i},
\end{align}
\end{small}
where $\mathbf{\xi}_{i}$ is the estimate error with the covariance matrix denoted as $\mathbf{J}_{i}$, related to the specific local solver. Due to the use of unbiased grid search, we have $\mathbb{E}\{\mathbf{\xi}_{i}\}=\mathbf{0}$, $Cov(\mathbf{s},\mathbf{\xi}_{i})=\mathbf{0}$.
Collecting all the distributed estimations yields a N-dimensional vector as follows
\begin{align}
&\hat{\mathbf{s}}=\mathbf{H}\mathbf{s}+\mathbf{\xi},~~\mathbf{H}\triangleq[\mathbf{I}_{3}, \mathbf{I}_{3}, \cdots, \mathbf{I}_{3}].
\end{align}
Under the  assumption that all UAV's estimation errors are independent with each other, the covariance matrix of $\mathbf{\xi}$ denoted as $\mathbf{J}$ is a  diagonal  matrix with the $i$-th diagonal element being $\mathbf{J_{i}}$.  According to the best linear unbiased estimation (BLUE) principle, the most reliable fusion implemented in the central UAV $1$  is given by
\begin{small}
\begin{align}\label{BLUE}
\hat{\mathbf{s}}_{c}&=(\mathbf{H}^{T}(\mathbf{J})^{-1}\mathbf{H})^{-1}\mathbf{H}^{T}(\mathbf{J})^{-1}\hat{\mathbf{s}}=\sum_{i=1}^{N}\underbrace{(\sum_{i=1}^{N}\mathbf{J}_{i}^{-1})^{-1}\mathbf{J}_{i}^{-1}}_{\mathbf{W}_{i}}\hat{\mathbf{s}}_{i}.\nonumber\\
\end{align}
\end{small}
The final estimation $\hat{\mathbf{s}}_{c}$ is performed in the central UAV. However, it is hard to compute the weight coefficients directly  while using complex algorithms in local estimation.A computable method is as follows
\begin{small}
\begin{align}\label{weightingmatrix}
\mathbf{W}_{i}&=(\sum_{i=1}^{N}\mathbf{J}_{i}^{-1})^{-1}\mathbf{J}_{i}^{-1}\approx(\sum_{i=1}^{N}\mathbf{C}_{i}^{-1})^{-1}\mathbf{C}_{i}^{-1}\nonumber\\
&=(\sum_{i=1}^{N}\mathbf{F}_{i})^{-1}\mathbf{F}_{i}\approx
(\sum_{i=1}^{N}\mathbf{F}_{i}(\hat{\mathbf{s}}_{i}))^{-1}\mathbf{F}_{i}(\hat{\mathbf{s}}_{i}),
\end{align}
\end{small}
where $\mathbf{C}_{i}$ and $\mathbf{F}_{i}$ represents for the Cramer-Rao lower bound (CRLB) and the Fisher information matrix (FIM) of the $i$-th UAV's estimation respectively.
The last term is gotten by replacing the unattainable true position of  target with the local estimation while calculating FIM.
It is reasonable if the accuracies of local estimations are acceptable. The mean-square error of the final fusion  in (\ref{weightingmatrix}) is
\begin{small}
\begin{align}
MSE&=\mathbb{E}\{||\hat{\mathbf{s}}_{c}- \hat{\mathbf{s}}||_2^2\}=\mathrm{Tr}(\sum_{i=1}^{N} \mathbf{W}_{i}^{T}\mathbf{J}_{i}\mathbf{W}_{i}) \nonumber\\
&\geq \underbrace{((\sum_{i=1}^{N}\mathbf{C}_{i}^{-1})^{-1})}_{MSE^{F}_{\min}} =\mathrm{Tr}(\mathbf{F}^{-1}),
\end{align}
\end{small}
where the $\mathbf{F}$ represents the FIM of the centralized UAV,,  fully utilizing all measurements from edge UAVs. The last term is from $\mathbf{F}\triangleq\sum_{i=1}^{N} \mathbf{F}_{i}$, as shown in (\ref{relationship}).

Finally, we make a comparison between it and other distributed estimation methods. The minimum MSE of the estimator  in \cite{4838824} (called DEM) denoted as $MSE^{M}_{\min}$,  is $(\sum_{i=1}^{N}(\mathrm{Tr}({\mathbf{C}_{i}))^{-1}})^{-1}$. Similarly, the minimum MSE of the average fusion (denoted as $MSE^{A}_{\min}$) is $\frac{1}{N}\mathrm{Tr}(\sum_{i=1}^{N}{\mathbf{C}_{i})}$. It is obvious that
\begin{small}
\begin{align}\label{compares}
 MSE^{F}_{\min}\leq MSE^{M}_{\min}\leq \min\limits_{i}({MSE_{\min}(i)})\leq MSE_{\min}^{A}.
\end{align}
\end{small}
The third term represents that the best lower bound of estimations among edge UAVs before fusion.
The second equality is achieved when $\mathbf{C}_{i}=\lambda\mathbf{I}$, where $\lambda$ is a constant. This condition is satisfied only when all edge UAVs are all in optimal geometry to  target (deduced from \cite{8698828 }). $\footnote{Let us give an example: for a fixed target, the equiangular structure geometry is optimal using RSS measurements\cite{8698828}.}$

\subsection{Communication overhead and Computational Complexity Analysis}
Now, let us make a complete analysis concerning communication overheads and computational complexities of  the proposed two methods: DMM and DEF combing local search with  DGN\cite{4291880}, DSCA$\footnote{The DSCA is a ditributed implementation of SCA after applying the nessary appriximation (\ref{approximation}).}$\cite{2018Parallel} and DEM\cite{4838824} combing local search. Here,  $\tau$  represents the dimension of the position vector, $p$ denotes the number of quantized bits of a signal element in the position vector and  $q$ denotes the number of bits of a single weighting coefficient. $k_{1}$ denotes the total iteration number of DMM.  Similarly, $k_{2}$ for  DGN and $k_{3}$ for DSCA.

The total numbers of required transmit bits, represented by $\mathcal{B}$, are given by
 \begin{small}
\begin{align}\label{CO-DMM}
\mathcal{B}_{DMM}=2(N-1)(\tau p)k_{1},
\end{align}
\begin{align}\label{CO-DGN}
\mathcal{B}_{DGN}=(N-1)(2\tau p+\tau^2 p)k_{2},
\end{align}
\begin{align}\label{CO-DSCA}
\mathcal{B}_{DSCA}=2(N-1)(\tau p)k_{3},
\end{align}
\begin{align}\label{CO-DEF}
\mathcal{B}_{DEF}=(N-1)(\tau p+\tau^2 q),
\end{align}
\begin{align}\label{CO-DEM}
\mathcal{B}_{DEM}=(N-1)(\tau p+q).
\end{align}
 \end{small}

The number of floating-point operations (FLOPs) of the above four methods are listed in Table~\ref{flops}$\footnote{For clarity and simplicity, we give the expression by ignoring all terms except for the leading (highest order or dominant) terms and let $M_{i}=M, i=1,2,..N$. It is notied that  $\tau$ is usually equal to $2$ or $3$,  thus the expression can be refined further. The final expression reflects how the computing cost varies as the number of UAVs and samples per UAV increase well.}$.

\begin{table}
\centering
\caption{Complexities of the proposed and comparison schemes (FLOPs)}
\resizebox{0.5\textwidth}{!}{
  \begin{tabular}{|c|c|c|c|c|}
\hline
  $\mathrm{Methods}$ & $\mathrm{Each~local~UAV}$ & $\mathrm{Central~UAV}$ &  $\mathrm{Total}$ &  $\mathrm{Simplification}$\\
  \hline
  Proposed DMM  & $O(3\tau M)k_{1}$  & $O(\tau N)k_{1}$&  $O(3\tau M N)k_{1}$   & $O(M N)k_{1}$ \\
  \hline
  DGN  & $O(2\tau^2M^{2})k_{2}$   &$O(\tau^2 N+2\tau^3)k_{2}$ &$O(2\tau^2 M^{2}N+\tau^2 N+2\tau^3)k_{2}$ & $O(M^{2}N)k_{2} $   \\
  \hline
   DSCA  & $O(3\tau M)k_{3}$   &$O(\tau N)k_{3}$ & $O(3\tau M N)k_{3}$ & $O(M N)k_{3}$   \\
  \hline
  Proposed DEF & $O(3\tau (\frac{l}{\Delta d})^{\tau}M)$  & $O(4 \tau^2 N+4\tau^3)$  &$O(3\tau (\frac{l}{\Delta d})^{\tau}MN+4 \tau^2 N+4\tau^3)$  &  $O((\frac{l}{\Delta d})^{\tau}MN) $ \\
  \hline
  DEM  & $O(3\tau (\frac{l}{\Delta d})^{\tau}M)$   & $O(3N)$  &$ O(3\tau (\frac{l}{\Delta d})^{\tau}MN) $  &  $O((\frac{l}{\Delta d})^{\tau}MN)$   \\
  \hline
\end{tabular}}
\label{flops}
\end{table}

\subsection{Derivation of CRLB}
In this section, we derive the CRLB of distributed multi-UAV  trajectory localization.
The FIM of the measurements from UAV $i$ and overall  measurements are denoted as  $\mathbf{F}_{i}$ and $\mathbf{F}$ respectively. The element in $a$-th row and $b$-th column of $\mathbf{F}_{i}$  is given by
 \begin{small}
\begin{align}
\mathbf{e}_{i,ab}=-\mathbb{E}\{\frac{\partial^2\log{Q_{i}(\mathbf{s})}}{\partial\phi_{a}\partial\phi_{b}}\}=\sigma_{i}^{-1}\sum_{j=1}^{M_{i}}\frac{\partial P_{i,j}}{\partial \phi_{a}}\frac{\partial P_{i,j}}{\partial \phi_{b}}, 1\leq a,b \leq 3,
\end{align}
 \end{small}
where
 \begin{small}
\begin{align}
[\frac{\partial P_{i,j}}{\partial \phi_{1}}, \frac{\partial P_{i,j}}{\partial \phi_{2}}, \frac{\partial P_{i,j}}{\partial \phi_{3}}]=\frac{-10\gamma_{i}}{\ln(10)}[\frac{x-x_{i,j}}{d_{i,j}^2}, \frac{y-y_{i,j}}{d_{i,j}^2}, \frac{z-z_{i,j}}{d_{i,j}^2}].
\end{align}
 \end{small}

The  element in the $a$-th row and the $b$-th column of $\mathbf{F}$ is given by
\begin{small}
\begin{align}
e_{ab}=-\mathbb{E}\{\frac{\partial^2\log{Q(\mathbf{s})}}{\partial\phi_{a}\partial\phi_{b}}\}=\sum_{i=1}^{N}\sigma_{i}^{-1}\sum_{j=1}^{M_{i}}\frac{\partial P_{i,j}}{\partial \phi_{a}}\frac{\partial P_{i,j}}{\partial \phi_{b}}=\sum_{i=1}^{N}e_{i,ab}.
\end{align}
\end{small}
The relationship between $\mathbf{F}$ and   $\mathbf{F}_{i}$ is obtained as follows
\begin{align}\label{relationship}
\mathbf{F}=\sum_{i=1}^{N} \mathbf{F}_{i}.
\end{align}

Finally, the CRLB of distributed multi-UAV  trajectory localization is expressed as
\begin{align}
\mathbf{C}=\frac{\mathrm{Tr}(\mathbf{F})^{-1}}{\det{\mathbf{F}}}=\frac{\sum_{a,b=1,a\neq b}^{3}(e_{aa}e_{bb}-e_{ab}^2)}{\tau},
\end{align}

where $\tau=e_{1,1}(e_{2,2}e_{3,3}-e_{2,3}^2)-e_{1,2}(e_{1,2}e_{3,3}-e_{1,3}e_{2,3})+ e_{1,3}(e_{1,2}e_{2,3}-e_{1,3}e_{2,2})$.

\section{Simulations and Discussions}

In this section, we evaluate the performance of the proposed distributed methods by  simulations. The target is  unknown in a $12\mathrm{km}\times 12\mathrm{km}$ AOI. System parameters are set as follows: $\gamma=3$, $\sigma^2=6\mathrm{dB}$, the number of samples per UAV is $8$. Each UAV flies over the  AOI with flight altitude $h=60\mathrm{m}$.

\begin{figure}
  \centering
  \includegraphics[width=0.45\textwidth]{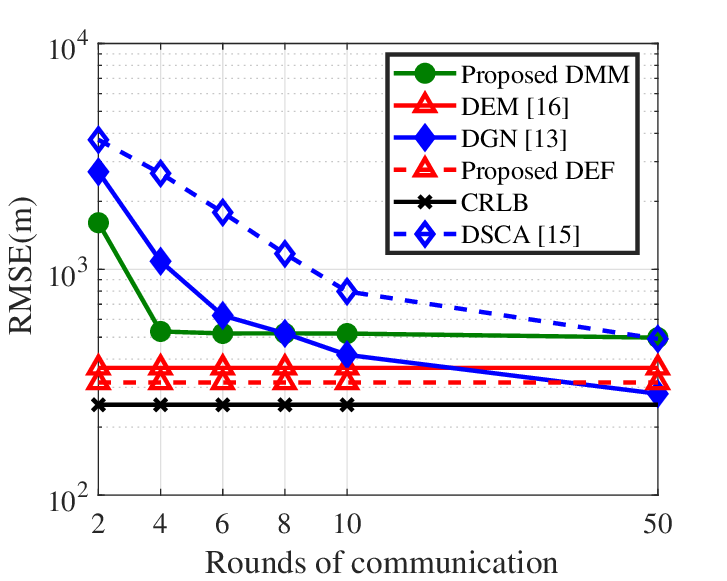}\\
  \caption{RMSE versus  rounds of communication between edge UAVs and the central UAV}  \label{roundsofcommunication}
\end{figure}

Fig.~\ref{roundsofcommunication} plots the curves of RMSE versus the rounds of communication between edge UAVs and the central UAV, namely iterations.  The $\Delta d$ of DEF/DEM is set to be $200\mathrm{m}$. As we can see,  the proposed DMM converges rapidly and has less estimated  error under low rounds of communication,  compared with DGN and DSCA. Besides, according to formulas about communication overhead,  the number of transmitted bits of DMM is same with DSCA and one third of DGN on each round. We can also see that the proposed DEF and DRM are most communication-effective, needing only one round of communication. However, according to Table.~ \ref{flops}, they have  higher computational complexity than the proposed DMM, DGN and DSCA.  Moreover, the proposed DEF is superior to the DEM in terms of RMSE performance.

\begin{figure}
  \centering
  \includegraphics[width=0.45\textwidth]{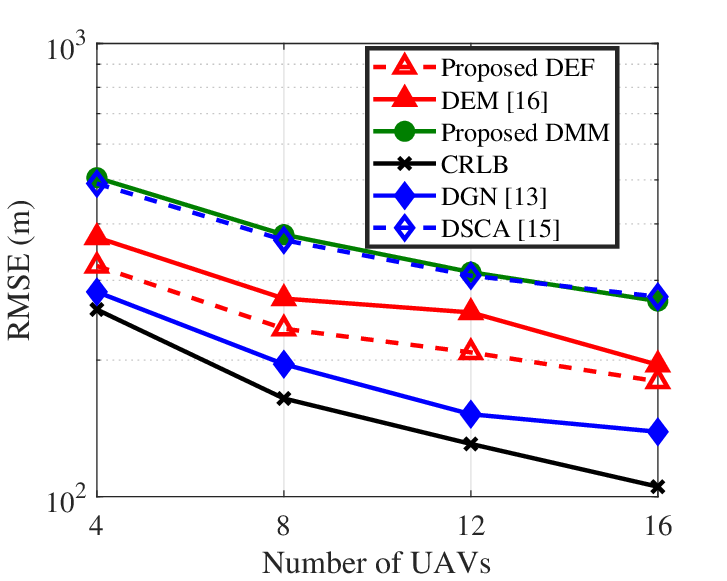}\\
  \caption{RMSE versus the number of UAVs}  \label{numberofUAVs}
\end{figure}

Fig.~\ref{numberofUAVs} demonstrates the curves of RMSE versus the number of  UAVs. As the number of UAVs increases, the RMSEs of all distributed methods increase gradually. They have an increasing order on performance as follows: proposed DMM/DSCA, DEM, proposed DEF, DGN. Thus, we can  make a conclusion that the proposed DMM strikes a good balance among performance, computational complexity and communication efficiency.

\section{Conclusion}
In our work, a RSS-based multi-UAV localization model utilizing trajectory has been established to find the position of passive source. To achieve considerable estimation under low communication overhead,  two high-communication-efficient distributed methods, DMM and DEF, have been proposed. The simulation results have shown that the proposed DMM has less estimated  error than the existing DGN and DSCA under low rounds of communication. Furthermore, the proposed DMM strikes a good balance among performance, computational complexity, and communication efficiency. The proposed DEF and DEM combing with local search are two highest communication-efficiency methods with higher computational complexity. Moreover, the proposed DEF outperforms DEM in terms of RMSE.
\appendices
\ifCLASSOPTIONcaptionsoff
  \newpage
\fi
\bibliographystyle{IEEEtran}
\bibliography{IEEEfull,cite}

\end{document}